# Causality Sum Rules in Conventional Scattering Matrices

Ning Han[1,2,3,†], Rui Zhao[1,2,†], Shuxing Yang[1,2,†], Mingzhu Li[4], Hongsheng Chen[1,2,5,*], Yihao Yang[1,2,5,*]

[1] Interdisciplinary Center for Quantum Information, State Key Laboratory of Extreme Photonics and Instrumentation, ZJU-Hangzhou Global Scientific and Technological Innovation Center, Zhejiang University, Hangzhou 310027, China

[2] International Joint Innovation Center, The Electromagnetics Academy at Zhejiang University, Zhejiang University, Haining 314400, China

[3] College of Optical and Electronic Technology, China Jiliang University, Hangzhou 310018, China

[4] School of Information and Electrical Engineering, Hangzhou City University, Hangzhou 310015, China

[5] Key Laboratory of Advanced Micro/Nano Electronic Devices & Smart Systems of Zhejiang, Jinhua Institute of Zhejiang University, Zhejiang University, Jinhua 321099, China

[†]These authors contributed equally to this work.

[*]Corresponding author. Email:
yangyihao@zju.edu.cn (Y. Y.);
hansomchen@zju.edu.cn (H. C.)

## Abstract

Scattering matrices are the standard experimental and computational description of photonic and electromagnetic devices. Passivity is explicit in the conventional incoming-outgoing matrix, whereas causality sum rules are usually formulated only after transforming the response into auxiliary variables. Here we show that these rules can be written directly in the conventional scattering matrix by removing the time advance introduced by the reference domain. Using the earliest-arrival delay of each channel, we define a domain-delayed matrix that preserves real-frequency passivity while restoring the causal time origin. Under explicit analyticity, transparency, and regularity assumptions, this matrix becomes a Schur function, enabling a Cayley-Herglotz construction. The resulting projected and determinant bounds constrain coherent channel superpositions and aggregate multichannel loss. The framework recovers Rozanov's absorber limit and spherical-multipole sum rules, while extending causality bounds to measurable quantities including insertion loss, suppressed singular-value channels, and conditional lossless delay-bandwidth trade-offs. Our work directly connects fundamental causality theory with experimentally accessible scattering data. The initial theoretical route is autonomously explored by Qiushi Engine, an AI research system for open-ended scientific discovery, and subsequently verified, refined, and developed by the authors, demonstrating a hybrid AI-human discovery workflow.

## Introduction

Scattering matrices are the working language of photonic and electromagnetic devices. They are what network analyzers measure, what full-wave solvers return, and what designers use to describe absorbers, antennas, metasurfaces, photonic switches, interferometer meshes, and phased arrays. A conventional incoming-outgoing matrix $S(\omega)$ at the operating frequency ω contains reflection, transmission, insertion loss, mode conversion, and coherent multiport interference in one experimentally accessible object. In this representation, passivity is almost immediate: for a power-normalized passive system, $S^{\dagger}(\omega)S(\omega) \preceq I$ on the real-frequency axis, where † is the conjugate transpose operator and $I$ is the unit matrix. Causality is not so directly visible. It is a statement about the temporal order of response, and in frequency space it appears through analyticity, dispersion relations, and sum rules that connect all frequencies. Thus, the matrix most natural for experiments and simulations makes energy conservation simple, but does not by itself display the full causality constraint.

This separation has a long history. In the 1920s, Kramers and Kronig related causal response to analyticity in the complex-frequency plane[1-3]. Around the same period and into the 1950s, Foster's reactance theorem and Bode-Fano matching theory showed how passivity and analyticity impose bandwidth limits on passive networks[4-6]. From the 1950s to the 1970s, bounded-real scattering matrices and $S$-matrix dispersion theory brought these ideas closer to network and scattering language[3,7]. Since the 2000s, electromagnetic sum rules have led to sharper physical limits, including Rozanov's absorber bound, spherical-multipole scattering bounds, antenna bandwidth limits, and more recent Green-function, local-conservation, and volume-$T$-operator bounds[8-23]. These developments established a broad principle: passive causal systems carry finite budgets in bandwidth, size, depth, loss, and delay. However, the causal bound formulations are usually obtained in alternative representations, such as Green functions, local conservation formulations, and volume $T$-operator approaches[21-23]. The missing link is a sum-rule formulation that acts directly on the conventional multi-channel scattering matrix $S(\omega)$ itself[24].

Here we show that causality sum rules can be written directly in the conventional scattering matrix once the reference-domain time advance is removed, thereby establishing a direct connection between experimentally accessible scattering measurements and fundamental causality limits. Our approach is based on separating the geometry-dependent reference contribution from the physical scattering response through a domain-delay correction. This transformation converts the conventional scattering matrix into a causal analytic object compatible with Schur-Cayley theory.

From this construction, we derive two universal causality sum rules: a projected bound for coherent channel superpositions and a basis-independent determinant bound for aggregate multichannel attenuation. These results recover established scalar limits, including Rozanov's absorber bound and spherical-multipole scattering bounds, while extending causality constraints to experimentally accessible multichannel quantities such as coherent suppression, insertion loss, and delay-bandwidth trade-offs. This connection enables causality constraints to be formulated directly in the experimentally accessible scattering representation.

Beyond the theoretical results, this work demonstrates a hybrid AI-human discovery workflow. Human researchers first defined the open-ended scientific question and research objective without prescribing a solution path. Qiushi Engine, an autonomous research system designed for sustained, thousand-step scientific reasoning[25], then conducted the exploration, developed derivations and numerical checks, and generated a traceable record of data, code, figures, research notes, and reports. The system identified the core theoretical route and produced the initial result, which the authors subsequently verified, refined into a rigorous theory, and physically interpreted, while retaining full responsibility for scientific validation and oversight.

**Main Text**

**Domain-delayed Schur construction**

We now construct the causal scattering representation required for the sum-rule formulation. The conventional scattering matrix contains a geometry-induced temporal advance arising from the finite scattering domain and the choice of channel reference surfaces. This reference-dependent contribution appears as a phase factor $e^{-i\omega T_a}$, where $T_a$ is the earliest-arrival delay operator, and obscures the causal analytic structure required for direct application of Schur-Cayley theory. To remove this reference contribution, we introduce the earliest-arrival delay operator $T_a = \mathrm{diag}(\tau_\alpha)_\alpha$, $\tau_\alpha \geq 0$, where $\tau_\alpha$ denotes the earliest arrival time of output channel $\alpha$. For spherical-wave scattering of an object enclosed by a sphere of radius $a$, all channels share the same delay, $\tau_\alpha = 2a/c$. The corresponding domain-delay operator is defined as $D_a = e^{i\omega T_a}$, and the domain-delayed scattering matrix is given by

$$\tilde{S}(\omega) = D_a(\omega)S(\omega). \tag{1}$$

Because $D_a(\omega)$ is unitary on the real-frequency axis, $D_a^\dagger(\omega)D_a(\omega) = I$, this transformation preserves the real-frequency power balance, $\tilde{S}^\dagger(\omega)\tilde{S}(\omega) = S^\dagger(\omega)S(\omega)$. Therefore, the passivity condition $S^\dagger(\omega)S(\omega) \preceq I$ is unchanged by the domain-delay correction. The transformation only removes the reference-dependent temporal advance and restores the causal analytic structure of the scattering response, as shown in Figs. 1a-c.

More precisely, the corrected scattering matrix $\tilde{S}(\omega)$ belongs to the class of operator Schur functions: it is analytic in the upper half-plane $\mathbb{C}^+$ and satisfies $\left\|\tilde{S}(\omega)\right\|_{op} \leq 1,\ z \in \mathbb{C}^+$. This is the central mathematical step of the construction. The construction assumes causal response, appropriate high-frequency transparency, and the regularity conditions required for the logarithmic Schur representation. The detailed proof is provided in Lemma S2.1 of the Supplementary Note 2. This result follows from the combination of causality, which provides the Hardy-class analytic continuation after the time-advance correction, and passivity, which preserves the contractive property on the real-frequency axis. The Schur property establishes the missing mathematical connection between experimentally accessible scattering data and causality-based sum rules. Applying the matrix Cayley transform, $W(\omega) = i[1+\tilde{S}(\omega)][1-\tilde{S}(\omega)]^{-1}$, maps the contractive scattering representation into an operator Herglotz function satisfying $\operatorname{Im} W(z) \succeq 0,\ z \in \mathbb{C}^+$. The Herglotz representation then converts the analyticity and passivity constraints of the corrected scattering matrix into a spectral integral relation, whose low-frequency expansion is determined by the geometric delay operator $T_a$. This relation provides the universal mechanism for extracting causality sum rules from experimentally accessible scattering responses. Notably, the entire route is accomplished through the cooperation of a hybrid AI-human discovery workflow, and the specific workflow is shown in Fig. 1d.

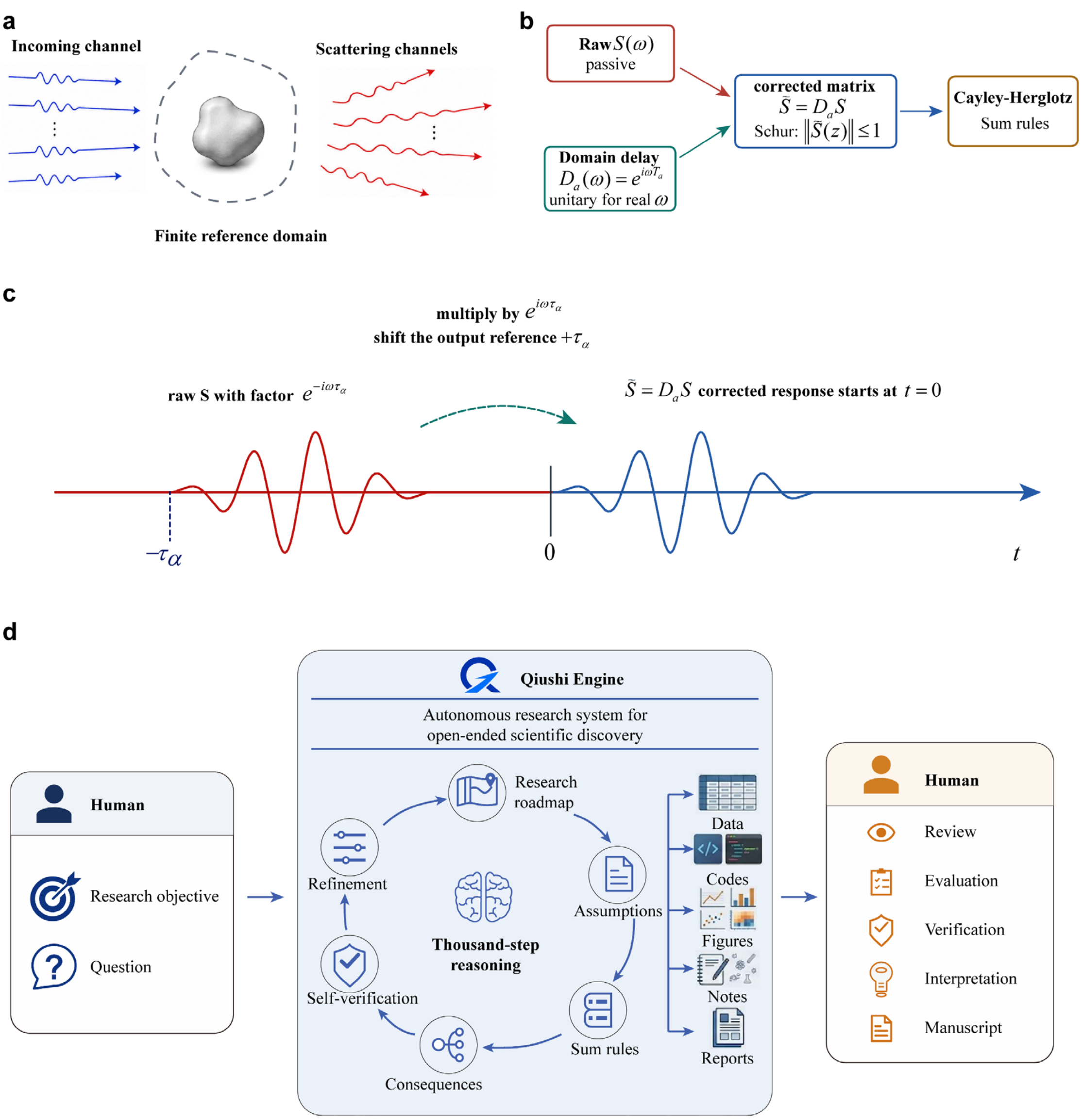

**Fig. 1 | Physical principle and AI-human hybrid scientific discovery workflow. a**, Finite reference domain defined by the scattering geometry and channel reference surfaces. **b**, Schur-Cayley-Herglotz machinery. **c**, Schematic of shifting the causal response to remove the time advance. A finite reference domain makes the conventional scattering matrix contain a channel-dependent apparent time-advance phase, even though the system is passive on the real-frequency axis. Multiplication by the domain-delay factor $D_a=e^{i\omega T_a}$ removes this reference phase without changing real-frequency power balance. The time axis indicates the raw apparent start at $t=-\tau_\alpha$ and the corrected causal start at $t=0$. Under the stated analyticity and transparency assumptions, the relative matrix $\tilde{S}=D_aS$ is the Schur object to which the Cayley-Herglotz and logarithmic sum-rule machinery applies. **d**, Hybrid human-AI scientific discovery workflow. Human scientists define open-ended research questions, while the Qiushi Engine proposes research roadmaps, explores candidate solutions, and generates extensive data, notes, reports, and code scripts. Subsequently, human scientists review, evaluate, and extend the generated outputs, and provide feedback when necessary.

For a delay eigenvector $v$ satisfying $T_a v = \tau_v v$, the scalar projection $\langle v, W(\omega)v \rangle$ inherits the Herglotz property. Its low-frequency expansion contains the causal delay contribution $\tau_v$, while the Herglotz integral representation relates this expansion coefficient to a weighted logarithmic integral of the scattering response. This yields the first projected causality sum rule.

**Projected sum rule**

The first result establishes a coherent-return suppression limit for channel superpositions within a common-delay subspace. Let $v$ be a unit incident superposition, and measure the component of the scattered field in the same superposition. The resulting amplitude is $s_v(\omega) = \langle v, S(\omega)v \rangle$ . To remove the domain phase without changing this scalar projection, $v$ must lie in an eigenspace of the delay operator, $T_a v = \tau_v v$. Then $\tilde{s}_v(\omega) = \langle v, \tilde{S}(\omega)v \rangle = e^{i\omega\tau_v} \langle v, S(\omega)v \rangle$. Consequently, $|\tilde{s}_v| = |s_v|$ on the real axis. If $v$ combines channels with different values of $\tau_\alpha$, the correction cannot be factored out as a single phase, and the scalar argument below does not apply. Thus, the positive Herglotz measure is set by the logarithmic suppression of the measured coherent return. Its first inverse-frequency moment is determined by the derivative of $h_v$ at the origin. See details in Supplementary Note 3.

Expanding at low frequency separates the geometric delay from the zero and phase-slope terms, convergence of the logarithmic integral and Blaschke product, and the angular-derivative condition, which ensures that the non-geometric terms do not increase the low-frequency coefficient: $hv' \le \tau_v = \langle v, T_a v \rangle$ . The Herglotz moment identity then yields

$$\left| \int_0^\infty \frac{\ln |\langle v, S(\omega)v \rangle|}{\omega^2} \mathrm{d}\omega \right| \le \frac{\pi}{2} \langle v, T_a v \rangle. \tag{2}$$

Equation (2) limits the bandwidth-integrated logarithmic suppression of return into the selected channel superposition $v$ by its domain-delay time $\tau_v$. Deep suppression over a broad band therefore requires a correspondingly large delay budget. For spherical waves, $T_a = (2a/c)I$ and the result holds for every superposition in the truncated channel space. For unequal channel delays, it holds within each common-delay eigenspace.

**Determinant sum rule**

The projected sum rule characterizes individual coherent scattering channels. To obtain a basis-independent constraint on the total multichannel response, we apply the same Schur-Cayley construction to the determinant of the domain-delayed scattering matrix $\det\tilde{S}$.

For a finite-dimensional $N$-channel system, the determinant magnitude satisfies $|\det S| = \prod_{j=1}^{N} \sigma_j(S)$, where $\sigma_j(S)$ are the singular values of $S$. Therefore, $-\ln|\det S| = \sum_j -\ln\sigma_j$ quantifies the aggregate logarithmic attenuation of the complete channel map. Unlike individual matrix elements or projected amplitudes, this quantity is invariant under unitary transformations of the input and output channel bases. Taking the determinant of the domain-delayed matrix gives $\det\tilde{S}(\omega) = \det D_a(\omega)\det S(\omega) = e^{i\omega \mathrm{tr} T_a}\det S(\omega)$. Because $\tilde{S}(\omega)$ is an operator Schur function, $\det\tilde{S}$ is a scalar Schur function. Applying the determinant form of the logarithmic Herglotz representation under the corresponding convergence, low-frequency unimodularity, and angular-derivative conditions yields the aggregated causality sum rule (See more details in Supplementary Note 4):

$$\left|\int_0^\infty \frac{\ln|\det S(\omega)|}{\omega^2}\mathrm{d}\omega\right| \le \frac{\pi}{2}\mathrm{tr} T_a. \tag{3}$$

The determinant bound provides a basis-independent constraint on the aggregate logarithmic attenuation of the multichannel scattering operator. While the projected bound describes suppression along a selected coherent channel, the determinant bound captures the collective contraction of all singular channels, including interchannel coupling and mode conversion. Therefore, it represents a genuinely multichannel causality constraint that cannot be obtained by independently applying scalar bounds to individual matrix elements.

**Physical consequences of the sum rules**

The projected and determinant sum rules lead to three directly measurable consequences: a depth-bandwidth constraint for a selected coherent return, an aggregate attenuation bound for the singular-value spectrum, and a bound on the number of independently suppressed singular channels, as shown in Fig. 2.

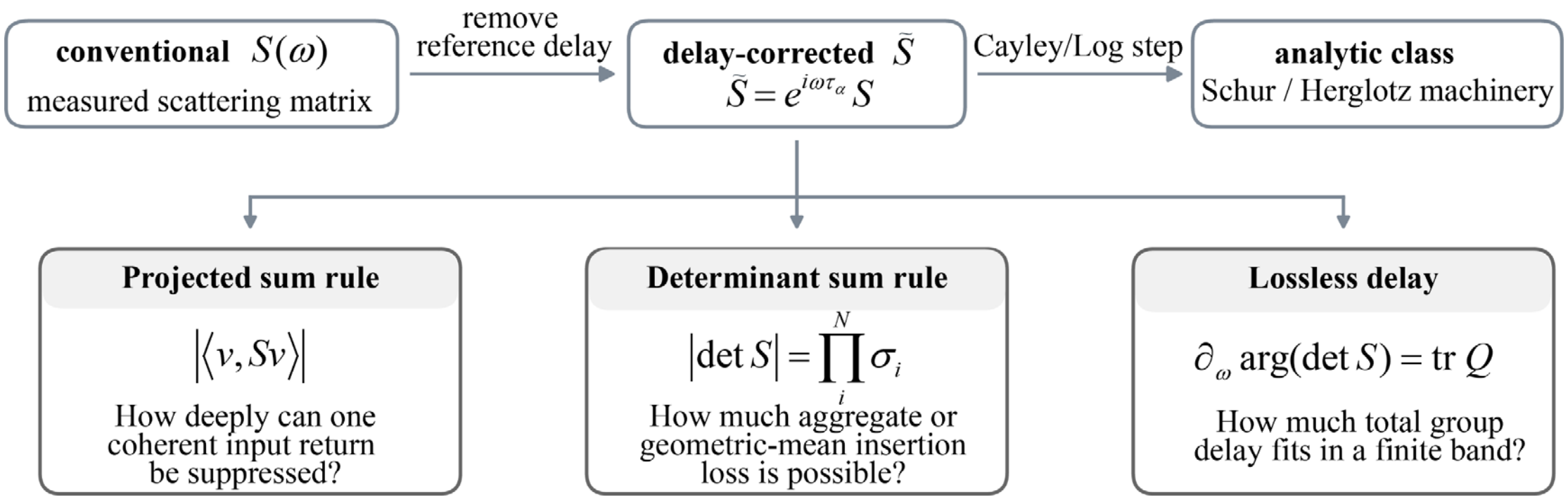

**Fig. 2 | From the conventional scattering matrix to observable bounds.** Removing the reference-domain advance gives the Schur matrix $\tilde{S} = D_a S$. A projection constrains coherent return, whereas the determinant constrains aggregate singular-value attenuation and suppressed-channel count. The determinant-phase branch is a separate conditional extension for lossless systems.

## From full-frequency constraints to finite-band bounds

The operator-level sum rules constrain the frequency-integrated logarithmic attenuation over the entire spectrum. To connect these universal constraints with experimentally relevant performance metrics, we convert the full-frequency integrals into finite-band depth-bandwidth relations.

Consider a dimensionless scattering observable $A(\omega)$ satisfying $0 < |A(\omega)| \le A_1 < 1$ for $\omega \in [\omega_0, \omega_0 + B]$. The contribution of this frequency interval to the logarithmic integral is bounded by $|\ln A_1| \int_{\omega_0}^{\omega_0+B} \omega^{-2} d\omega$. For a narrow fractional bandwidth $\beta = B/\omega_0 \ll 1$, this contribution can be approximated as $|\ln A_1| B/\omega_0^2$.

Therefore, the full-frequency sum rules impose finite-band constraints on the achievable attenuation depth, bandwidth, and physical size. For the projected sum rule, the relevant quantity is the coherent return amplitude $s_v(\omega) = \langle v, S(\omega) v \rangle$. If the coherent response remains below $|s_v(\omega)| \le s_0$ over a fractional bandwidth $\beta = B/\omega_0$, the integral constraint gives a depth-bandwidth limitation determined by the available delay budget $\langle v, T_a v \rangle$.

For the determinant sum rule, the corresponding quantity is the aggregate channel response $|\det S(\omega)|$. If $|\det S(\omega)| \le \Delta_0$ within the operating band, the finite-band constraint becomes a limit on the total multichannel attenuation budget determined by $\mathrm{tr}\, T_a$.

The narrowband approximation $\beta \ll 1$ is used for the expressions and figures below. The exact conversion between the full-frequency integral constraints and finite-band performance metrics, including finite-band corrections beyond the narrowband limit, is provided in Supplementary Note 5.

**Consistency checks: scalar limits.**

The two operator-level sum rules recover several established scalar causality bounds as limiting cases. These reductions demonstrate that the domain-delayed Schur-Cayley construction provides a unified scattering-matrix formulation of previously known fundamental limits.

First, consider a single-channel passive system, where the scattering matrix reduces to a scalar reflection coefficient, $S(\omega)=\Gamma(\omega)$. For a planar absorber of thickness $d$ backed by a perfect conductor, the earliest round-trip arrival time is $\tau_v = 2d/c$. Substituting this into the projected sum rule [see Eq. (2)], gives $\int_0^\infty -\ln|\Gamma(\omega)|/\omega^2 \mathrm{d}\omega \le \pi\tau_v/2 = \pi d/c$. This directly recovers the Rozanov absorber bound[8]. In the narrowband approximation, where the reflection magnitude is approximately constant within a bandwidth $B$ centered at $\omega_0$, the above inequality becomes $|\ln R_0| B/\omega_0 \le \pi d/\lambda_0$, where $R_0 = |\Gamma(\omega_0)|$, $\lambda_0 = 2\pi c/\omega_0$. Thus, the conventional thickness-bandwidth trade-off of absorbers emerges as a special case of the general projected sum rule.

A second recovery follows for spherical-wave scattering. Consider a passive scatterer enclosed by a sphere of radius $a$. For spherical multipole channels, the earliest arrival delay is identical for all channels, $T_a = (2a/c)I$. Therefore, every normalized channel superposition satisfies the delay-eigenvector condition, $T_a v = (2a/c)v$. Choosing a single spherical multipole basis vector, $v = e_n^{(\sigma)}$, the projected sum rule reduces to $\int_0^\infty -\ln|s_n^{(\sigma)}(\omega)|/\omega^2 \mathrm{d}\omega \le \pi a/c$, which exactly recovers the Bernland-Gustafsson spherical-multipole scattering bound[9,10]. However, unlike the original diagonal formulation, the present result extends this constraint from individual multipole channels to arbitrary coherent superpositions within a common-delay eigenspace.

These recoveries confirm that the known scalar limits are embedded within the general $S$-matrix formulation. The remaining advantage of the operator-level theory is its ability to constrain collective scattering behavior beyond individual coefficients. We next explore these multichannel consequences, beginning with coherent return suppression in common-delay channel subspaces.

**Consequence 1: coherent-return suppression.**

The projected sum rule establishes a fundamental limit on coherent return suppression for channel superpositions within a common-delay subspace, extending conventional bounds beyond individual diagonal scattering coefficients to coherent combinations of channels.

For a normalized channel superposition $v$, the coherent return amplitude is $s_v = \langle v, S(\omega)v \rangle$. When $v$ is a delay eigenvector satisfying $T_a v = \tau_v v$, the projected sum rule gives $\left| \int_0^{\infty} -\ln|s_v(\omega)|/\omega^2 \mathrm{d}\omega \right| \le \pi\tau_v/2$. which directly relates the achievable broadband coherent suppression to the available causal delay budget of the scattering structure.

To illustrate the physical meaning, consider spherical-wave scattering from an object enclosed by a sphere of radius $a$. In this case, $T_a = (2a/c)I$, so every coherent superposition of spherical channels satisfies the delay-eigenvector condition with $\tau_v = 2a/c$. Therefore, the suppression of any coherent multipole superposition is fundamentally limited by the physical size of the scatterer:

$$\left| \int_0^{\infty} \frac{-\ln|\langle v, S(\omega)v \rangle|}{\omega^2} \mathrm{d}\omega \right| \le \pi a/c. \tag{4}$$

This result shows that the same causal delay budget constrains not only individual multipole channels but also coherent combinations of channels.

For a coherent return satisfying $|s_v(\omega)| \le \rho_v$ over a fractional bandwidth $\beta = B/\omega_0$, define the coherent suppression depth as $D_v = 20\log_{10}(1/\rho_v)$, Using the narrowband approximation, the bound becomes $D_v \le (20\pi/\ln 10)k_0 a/\beta$.

Figure 3a shows this depth-bandwidth-size boundary for different electrical sizes $k_0 a$. The suppression depths below each curve satisfy the causal sum-rule constraint. For example, a 60-dB return suppression over a 10% fractional bandwidth corresponds to $\rho_v = 10^{-3}$ and requires $k_0 a \gtrsim 0.22$. This is a necessary causality condition rather than a sufficient design criterion.

**Consequence 2: aggregate and geometric-mean attenuation.**

The coherent-return suppression limit characterizes attenuation along a selected coherent channel superposition. By contrast, the determinant sum rule provides a complementary constraint on the collective attenuation of the full multichannel scattering operator.

For an $N$-channel scattering system, $|\det S| = \prod_{j=1}^{N} \sigma_j[S(\omega)]$, where $\sigma_j(S)$ are the singular values of the scattering matrix. Therefore, $-\ln|\det S| = \sum_j -\ln \sigma_j$, measures the aggregate logarithmic attenuation accumulated over all singular scattering channels.

Unlike individual matrix elements or projected coherent amplitudes, the determinant is invariant under unitary transformations of the input and output channel bases. Therefore, it provides a basis-independent measure of the collective contraction of the scattering operator.

The determinant constrains aggregate logarithmic attenuation across the included singular channels, rather than the attenuation of an individual multipole or matrix element. This quantity can be interpreted as absorption only when the retained scattering channels contain all power-carrying output channels. Otherwise, a reduced determinant magnitude may also arise from radiation or coupling into omitted channels.

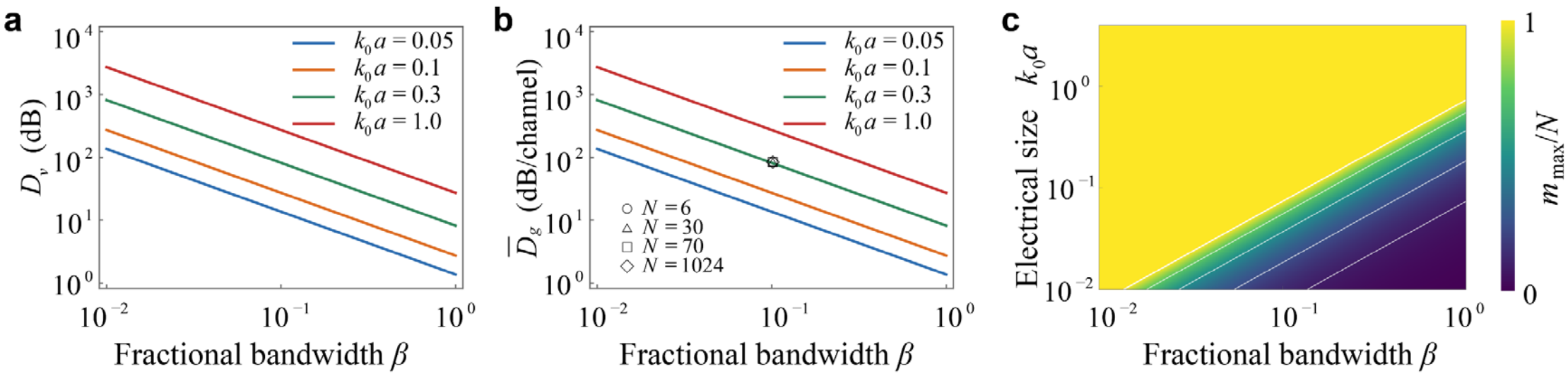

**Fig. 3 | Finite-band consequences of the projected and determinant sum rules. a**, Maximum coherent-return depth $D_v = 20\log_{10}(1/\rho_v)$ for a prescribed fractional bandwidth and electrical size. **b**, Upper bound on the geometric-mean singular-value depth after normalizing the determinant result by the number of channels; this upper bound is independent of $N$. **c**, Maximum fraction $m_{max}/N$ of singular channels that can remain below the prescribed threshold $\rho_\sigma$ across the operating bandwidth.

For a spherical-wave basis truncated at multipole order $n = N_{max}$, accounting for the $2n+1$ azimuthal orders and two polarizations for each multipole order, the finite truncation contains $N = 2N_{max}(N_{max}+2)$ open scattering channels. For a spherical reference domain, $T_a = (2a/c)I$, and therefore $\mathrm{tr}\, Ta = 2aN/c$. Applying the determinant sum rule gives the aggregate attenuation bound

$$\int_0^\infty \frac{-\ln|\det S(\omega)|}{\omega^2} \mathrm{d}\omega \le \frac{N\pi a}{c}. \tag{5}$$

To obtain a finite-band form, define the determinant modulus within an operating band as $\Delta_0 = \sup_{\omega\in[\omega_0,\omega_0+B]} |\det S(\omega)|$. Using the narrowband approximation, the determinant constraint becomes $|\ln \Delta_0| B/\omega_0 \pi \le N k_0 a$. Equivalently, defining the aggregate determinant suppression depth $D_{\det} = 20\log_{10}(1/\Delta_0)$, we obtain $D_{\det} B/\omega_0 \le 27.3 N k_0 a$ (See more details in Supplementary Note 6).

The total determinant budget grows linearly with the number of channels $N$, but the number of singular values sharing this budget also grows linearly. Therefore, after normalization by the channel number, the geometric-mean attenuation exhibits an $N$-independent causal limit. Define the geometric-mean singular value $\bar{\sigma}_g = |\det S|^{1/N}$ and its depth $\bar{D}_g = -20\log_{10}\bar{\sigma}_g$. Dividing the determinant constraint by $N$ gives an upper bound on the geometric-mean singular-value depth $\bar{D}_g B/\omega_0 \le (20\pi/\ln 10) k_0 a$, which is independent of the number of channels, as shown in Fig. 3b. This is an average constraint and does not require equal attenuation across all singular channels; different channels may experience different suppression levels while the geometric mean remains bounded by the available causal delay budget. For $N = 70$, a determinant depth of 80 dB corresponds to a geometric-mean depth of $80/70 \simeq 1.1$ dB. This does not imply equal suppression of every singular channel.

**Consequence 3: suppressed-channel count.**

The geometric-mean attenuation bound constrains the average suppression of the singular-value spectrum, but it does not determine how many independent scattering channels can simultaneously achieve a prescribed attenuation level. We therefore derive a bound on the number of suppressed singular channels.

Choose a suppression threshold $0 < \rho_\sigma < 1$. If at least $m$ singular values satisfy $\sigma_j(S) \le \rho_\sigma$ throughout an operating band, then passivity gives $|\det S| \le \rho_\sigma^m$. Combining this inequality with the determinant sum rule and the finite-band approximation yields $m \le \pi N k_0 a / (|\ln \rho_0| B/\omega_0)$, together with $m \le N$, this gives the maximum number of singular channels that can simultaneously maintain a target suppression level over a finite bandwidth (See more details in Supplementary Note 7). Here $m$ counts orthogonal incident singular vectors, not physical ports. Figure 3c plots the maximum fraction $m/N$ for $\rho_\sigma = 0.1$.

When the scattering matrix contains all power-carrying exterior output channels, the singular values can be related to absorption eigenchannels through $A_j = 1 - \sigma_j^2$. Requiring at least $m$ eigenchannels to satisfy $A_j \geq A_0$ throughout the operating band is equivalent to $\rho_\sigma = \sqrt{1 - A_0}$ and gives $m \leq 2\pi N k_0 a / (|\ln(1 - A_0| B / \omega_0)$. For a channel-complete $N = 30$ model, maintaining $A_j \geq 0.99$ in all 30 eigenchannels over a 10% fractional bandwidth requires $k_0 a \geq 7.3 \times 10^{-2}$. This is a necessary condition from the determinant budget, rather than a sufficient condition for constructing such a device.

If the retained scattering channels do not include all power-carrying output channels, a small singular value indicates only that power has left the measured subspace. It does not distinguish absorption from radiation or coupling into unmeasured channels. Therefore, the absorption interpretation requires a channel-complete scattering description or a closed reducing subspace with no coupling to omitted channels.

**Conditional phase-delay extension for lossless multiport systems**

The preceding consequences follow from the projected and determinant logarithmic sum rules and constrain attenuation-related observables. Lossless systems represent a distinct regime: because $|\det S| = 1$, on the real-frequency axis, the logarithmic attenuation bounds become trivial. In this case, the relevant causal quantity is the accumulated scattering phase rather than the modulus.

For a lossless multiport system, $S$ is unitary and $|\det S| = 1$, the determinant can be expressed as $\det S(\omega) = e^{i\Theta(\omega)}$. The Wigner-Smith time-delay operator is defined as $Q(\omega) = -iS^\dagger \partial_\omega S$, which gives the phase-delay identity $\Theta'(\omega) = \mathrm{tr}Q(\omega)$. Therefore, $\mathrm{tr}\,Q(\omega)$ represents the total delay accumulated over all scattering channels.

Define the band-averaged delay trace as $\overline{\mathrm{tr}Q} = \frac{1}{B}\int_{\omega_0}^{\omega_0+B} \mathrm{tr}Q(\omega)\mathrm{d}\omega$. Unlike the attenuation bounds, obtaining a universal delay-bandwidth estimate requires an additional spectral-counting assumption. Let $\ell$ denote the effective propagation length entering the modal-count model, and $\lambda_0 = 2\pi c / \omega_0$. If the delay-corrected determinant contains no singular inner factor and its Blaschke phase accumulation satisfies the modal-count hypothesis $H_B$ (see details in Supplementary Note 8), then

$$\overline{\mathrm{tr}Q} B \leq 4\pi^2 N \frac{\ell}{\lambda_0}. \tag{6}$$

By using the fractional bandwidth $\beta = B / \omega_0$, the delay-bandwidth relation can be written as $\overline{\mathrm{tr}Q}\omega_0 \leq 4\pi^2 N(\ell / \lambda_0) / \beta$. For the mean delay per channel $q = \overline{\mathrm{tr}Q} / N$, one obtains the basis-independent form $qB \le 4\pi^2 \ell / \lambda_0$ (See more details in Supplementary Notes 8 and 9).

For example, consider a target mean normalized delay of $q\omega_0 = 10^6$ over a fractional bandwidth 1%. The required delay-bandwidth product is then $qB = 10^4$. Under the modal-count hypothesis, this requires $\ell / \lambda_0 \gtrsim 2.5 \times 10^2$. Thus, achieving extremely large delay over finite bandwidth necessarily requires a correspondingly large effective propagation length. For an $N$-channel implementation, the aggregate delay trace scales as $\overline{\mathrm{tr}Q} = Nq$.

Figure 4a illustrates the linear scaling of the aggregate delay budget with the number of channels $N$ at fixed $\ell / \lambda_0$, while Fig. 4b shows the reciprocal trade-off between delay and fractional bandwidth. These relations are relevant to delay lines, beamforming networks, and interferometric photonic architectures.

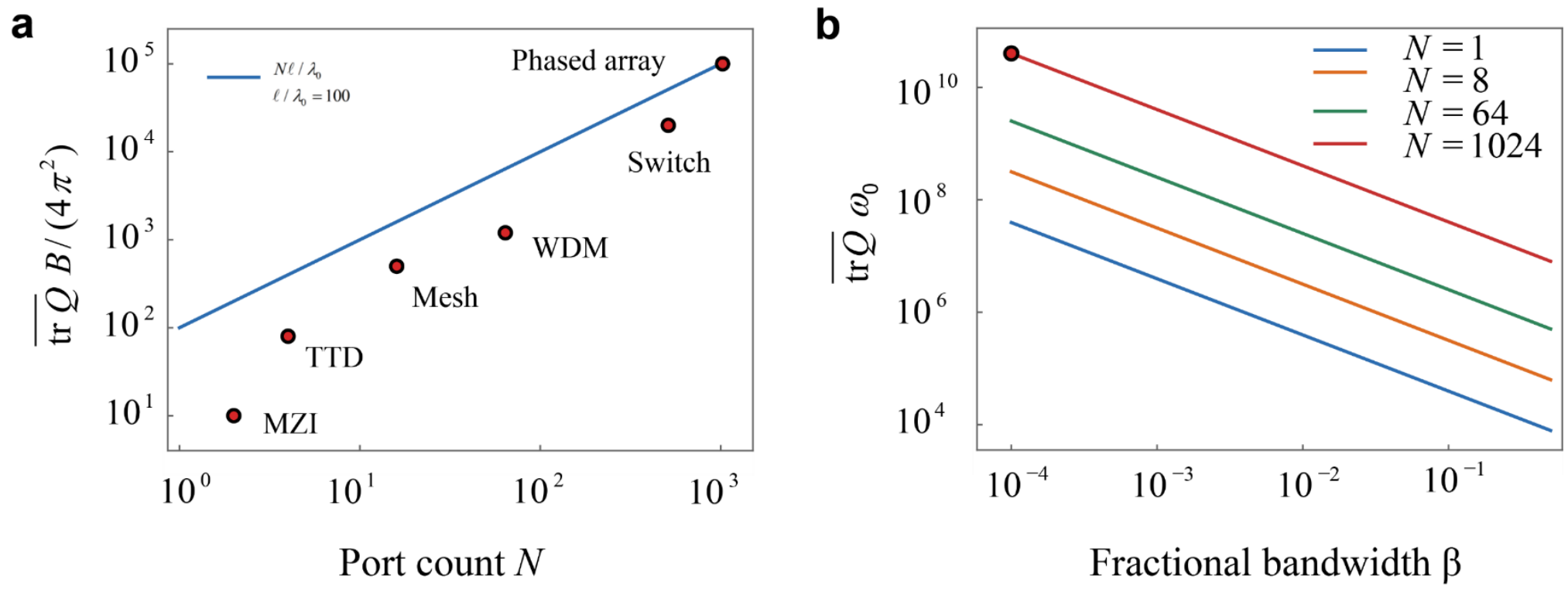

**Fig. 4 | Conditional delay-bandwidth estimates**. **a**, Aggregate delay-bandwidth envelope at $\ell / \lambda_0 = 100$ and illustrative network scales. **b**, Equivalent bound on $\overline{\mathrm{tr}}Q\omega_0$ as a function of fractional bandwidth. Additional high-$Q$ phase winding is not included.

Importantly, the delay-bandwidth estimate relies on the modal-count hypothesis $H_B$. Sharp resonances can introduce additional Blaschke phase accumulation beyond the assumed spectral count. An additional phase contribution $k_{\mathrm{extra}}$ modifies the bound by an additive term $2\pi k_{\mathrm{extra}}$ in the integrated trace delay. Therefore, Eq. (6) should be interpreted as conditional spectral-counting estimates rather than universal limits applicable to arbitrary high-$Q$ resonant or slow-light systems.

*On the use of Qiushi Engine*. Because there is growing interest in whether autonomous AI systems can contribute to real scientific discovery, we state explicitly how Qiushi Engine entered this work. The path to the present theory also illustrates a new mode of scientific discovery. Qiushi Engine participated as an autonomous exploration partner within a hybrid human-AI discovery workflow. Given the open problem of formulating causality sum rules directly in the conventional scattering matrix, the system explored candidate mathematical routes, identified the domain-delay correction strategy, generated preliminary derivations, and organized intermediate research artifacts. These outputs were not treated as proofs. The authors subsequently audited the analytic assumptions, corrected the functional-analytic formulation, verified scalar recoveries, examined numerical consequences, and established the final physical interpretation. This division of roles demonstrates a research paradigm in which AI expands the space of candidate scientific pathways, while human researchers retain rigorous validation and scientific responsibility. Supplementary Note 10 records this workflow.

**Discussion**

The domain-delayed Schur-Cayley-Herglotz construction provides a direct route to formulating causality sum rules in the conventional scattering matrix. Removing the apparent time advance introduced by the finite reference domain restores the analytic structure required for the sum-rule construction. This leads to two operator-level results: a projected sum rule for coherent channel superpositions and a basis-independent determinant sum rule for aggregate multichannel attenuation. The framework recovers established scalar limits, including Rozanov's absorber bound and spherical-multipole scattering sum rules, while extending them to measurable multichannel quantities such as coherent suppression, determinant insertion loss, singular-channel suppression, and lossless delay.

Working in the conventional scattering representation keeps these bounds directly connected to the input-output data obtained from experiments, full-wave simulations, and inverse design. The framework complements approaches based on polarizabilities, Green functions, and volume ($T$)-operators[21-23], which express causality through material or source variables, by providing an equivalent formulation at the scattering level. The projected rule constrains a selected coherent channel superposition, whereas the determinant rule captures the collective attenuation of the full multichannel response.

More broadly, this work connects fundamental causality theory with the scattering data routinely used in photonics and electromagnetism. Qiushi Engine contributed to the initial exploration of the theoretical route, including candidate formulations, preliminary derivations, numerical checks, and research records. The authors then verified the assumptions, completed the derivation, and established the final physical interpretation. This workflow provides a concrete example of how autonomous research systems can support theoretical discovery while rigorous validation and scientific judgement remain with the researchers.

## Methods

### Scattering convention and channel normalization

We use the time dependence $\exp(-i\omega t)$. Incoming and outgoing channel amplitudes are collected in vectors $a(\omega)$ and $b(\omega)$ and related by

$$b(\omega) = S(\omega)a(\omega).$$

The channels are power normalized, so that $\|a\|^2$ and $\|b\|^2$ represent incident and outgoing powers. Passivity therefore gives

$$S^\dagger(\omega)S(\omega) \preceq I,$$

on the real-frequency axis. Determinants are taken only after restricting the channel map to a finite set of propagating channels. When attenuation is interpreted as absorption, this set must include all propagating output channels into which the chosen inputs can scatter. Otherwise, the missing power may represent radiation or mode conversion into omitted channels rather than dissipation.

### Domain-delay correction

The conventional scattering matrix depends on the reference surfaces and on the finite domain used to define the incoming and outgoing channels. For output channel $\alpha$, let $\tau_\alpha \geq 0$ denote the magnitude of the earliest apparent time advance introduced by this convention. The correction times are assembled into

$$T_a = \mathrm{diag}(\tau_\alpha)_\alpha,$$

and the domain-delay operator is

$$D_a(\omega) = \exp(i\omega T_a) = \mathrm{diag}\left(\exp(i\omega\tau_\alpha)\right)_\alpha.$$

We then define the corrected scattering matrix

$$\tilde{S}(\omega) = D_a(\omega)S(\omega).$$

Because $D_a$ is unitary for real $\omega$, this transformation preserves the real-frequency power balance. In the time domain, if the uncorrected impulse response $s_{\alpha\beta}(t)$ can begin at $t = -\tau_\alpha$, multiplication by $\exp(i\omega\tau_\alpha)$ gives

$$\tilde{s}_{\alpha\beta}(t) = s_{\alpha\beta}(t - \tau_\alpha),$$

whose support begins at $t = 0$. Thus, $\tau_\alpha$ is a reference-domain correction time. It is not a resonance lifetime, a Wigner-Smith group delay, or the delay of an output-pulse maximum. For a spherical reference surface of radius $a$, the open spherical-wave channels have $\tau_\alpha = 2a/c$.

**Acknowledgments**

The work at Zhejiang University sponsored by the Key Research and Development Program of the Ministry of Science and Technology under Grants No. 2022YFA1405200 (Y.Y.), No. 2022YFA1404900 (Y.Y.), and No 2022YFA1404704 (H.C.), the Fundamental Research Funds for the Zhejiang Provincial Universities No. 226-2025-00231 (Y.Y.), the Science Challenge Project No. TZ2025015 (Y.Y.), the National Natural Science Foundation of China No. U25D8017 (Y.Y.), the National Natural Science Foundation of China (NSFC) under grant No. 61975176 (H.C.), the Key Research and development Program of Zhejiang Province under grant no. 2022C01036 (H.C.), the National Natural Science Foundation of China (NSFC) under Grant No. 12504497 (N.H.), the National Natural Science Foundation of China (NSFC) under Grant No. 12304472 (M.L.), the Baima Lake Laboratory Joint Fund of the Zhejiang Provincial Natural Science Foundation of China under Grants No. LBMHY25A040002 (M.L.), the Hangzhou Natural Science Foundation under Grants No. 2024SZRYBF050004 (M.L.), and the Natural Science Foundation of Zhejiang Province under Grant No. LQN26A040006 (N.H.).

**Author contributions**

Y.Y. conceived the project; N.H., R.Z., S.Y. and M.L. verified the assumptions, limiting cases and numerical examples; H.C. and Y.Y. supervised the project; N.H. and R.Z. wrote the manuscript with input from S.Y., M.L., H.C. and Y.Y. All authors discussed the results and reviewed the manuscript.

**Competing interests**

The authors declare no competing interests.

**Data and materials availability**

All data that support the findings of this study are present in the paper and the Supplementary Information. Any additional information may be available from the corresponding authors upon request.

## References

1. Kramers, H. A. La diffusion de la lumière par les atomes. *Atti Congr. Internazionale Dei Fis.* **2**, 545-557 (1927).
2. Kronig, R. de L. On the theory of dispersion of X-rays. *J. Opt. Soc. Am.* **12**, 547-557 (1926).
3. Nussenzveig, H. M. *Causality and Dispersion Relations*. (Academic Press, New York, 1972).
4. Foster, R. M. A reactance theorem. *Bell Syst. Tech. J.* **3**, 259-267 (1924).
5. Bode, H. W. *Network Analysis and Feedback Amplifier Design*. (Van Nostrand, New York, 1945).
6. Fano, R. M. Theoretical limitations on the broadband matching of arbitrary impedances. *J. Frankl. Inst.* **249**, 57-83 (1950).
7. Youla, D. C., Castriota, L. J. & Carlin, H. J. Bounded real scattering matrices and the foundations of linear passive network theory. *IRE Trans. Circuit Theory* **6**, 102-124 (1959).
8. Rozanov, K. N. Ultimate thickness to bandwidth ratio of radar absorbers. *IEEE Trans. Antennas Propag.* **48**, 1230-1234 (2000).
9. Bernland, A., Gustafsson, M. & Nordebo, S. Physical limitations on the scattering of electromagnetic vector spherical waves. *J. Phys. Math. Theor.* **44**, 145401 (2011).
10. Bernland, A. Bandwidth limitations for scattering of higher order electromagnetic spherical waves with implications for the antenna scattering matrix. *IEEE Trans. Antennas Propag.* **60**, 4345-4353 (2012).
11. Bernland, A., Luger, A. & Gustafsson, M. Sum rules and constraints on passive systems. *J. Phys. Math. Theor.* **44**, 145205 (2011).

12. Sohl, C., Gustafsson, M. & Kristensson, G. Physical limitations on broadband scattering by heterogeneous obstacles. *J. Phys. Math. Theor.* **40**, 11165-11182 (2007).
13. Sohl, C., Larsson, C., Gustafsson, M. & Kristensson, G. A scattering and absorption identity for metamaterials: experimental results and comparison with theory. *J. Appl. Phys.* **103**, 054906 (2008).
14. Gustafsson, M., Sohl, C., Larsson, C. & Sjöberg, D. Physical bounds on the all-spectrum transmission through periodic arrays. *EPL* **87**, 34002 (2009).
15. Biehs, S.-A., Rousseau, E. & Greffet, J.-J. Mesoscopic description of radiative heat transfer at the nanoscale. *Phys. Rev. Lett.* **105**, 234301 (2010).
16. Gustafsson, M., Vakili, I., Keskin, S. E. B., Sjöberg, D. & Larsson, C. Optical theorem and forward scattering sum rule for periodic structures. *IEEE Trans. Antennas Propag.* **60**, 3818-3826 (2012).
17. Miller, O. D., Johnson, S. G. & Rodriguez, A. W. Shape-independent limits to near-field radiative heat transfer. *Phys. Rev. Lett.* **115**, 204302 (2015).
18. Shim, H., Fan, L., Johnson, S. G. & Miller, O. D. Fundamental limits to near-field optical response over any bandwidth. *Phys. Rev. X* **9**, 011043 (2019).
19. Molesky, S., Jin, W., Venkataram, P. S. & Rodriguez, A. W. T-operator bounds on angle-integrated absorption and thermal radiation for arbitrary objects. *Phys. Rev. Lett.* **123**, 257401 (2019).
20. Kuang, Z. & Miller, O. D. Computational bounds to light–matter interactions via local conservation laws. *Phys. Rev. Lett.* **125**, 263607 (2020).
21. Molesky, S., Chao, P., Jin, W. & Rodriguez, A. W. Global T operator bounds on electromagnetic scattering: upper bounds on far-field cross sections. *Phys. Rev. Res.* **2**, 033172 (2020).

22. Molesky, S., Chao, P., Mohajan, J., Reinhart, W., Chi, H. & Rodriguez, A. W. T-operator limits on optical communication: metaoptics, computation, and input-output transformations. *Phys. Rev. Res.* **4**, 013020 (2022).
23. Zhang, L., Monticone, F. & Miller, O. D. All electromagnetic scattering bodies are matrix-valued oscillators. *Nat. Commun.* **14**, 7724 (2023).
24. O. D. Miller and F. Monticone, Fundamental limits in photonics and electromagnetics: a tutorial, arXiv:2605.24738, 2026.
25. Yang, S., et al. End-to-end autonomous scientific discovery on a real optical platform, arXiv:2604.27092 (2026).